\documentclass[prl,twocolumn,floatfix,superscriptaddress]{revtex4}
\usepackage{amssymb}
\usepackage{amsmath}
\usepackage{graphicx}
\usepackage{hyperref}
\usepackage{multirow}
\usepackage{array}
\usepackage{float}

\def\rd{R_{\mathrm{D}}}
\def\rgd{\delta R_{\mathrm{D}}}
\def\vd{V_{\mathrm{D}}}
\def\vh{V_{\mathrm{H}}}
\def\um{\mathrm{\mu m}}

\def\perm{\mathrm{m}^{-2}}
\def\mob{\mathrm{m}^{2}}

\def\vg{V_{\mathrm{g}}}
\def\vqpc{\nu_{\mathrm{c}}}

\def\rh{R_{\mathrm{H}}}

\begin{document}
\title{Coulomb Oscillations in Antidots in the Integer and Fractional Quantum Hall Regimes}
\author{A.~Kou}
\affiliation{Department of Physics, Harvard University, Cambridge, Massachusetts
02138, USA}

\author{C.~M.~Marcus}
\affiliation{Department of Physics, Harvard University, Cambridge, Massachusetts
02138, USA}

\author{L.~N.~Pfeiffer}
\affiliation{Department of Electrical Engineering, Princeton University, Princeton, New
Jersey 08544, USA}

\author{K.~W.~West}
\affiliation{Department of Electrical Engineering, Princeton University, Princeton, New
Jersey 08544, USA}
\date{\today}

\begin{abstract}
We report  measurements of resistance oscillations in micron-scale antidots in both the integer and fractional quantum Hall regimes. In the integer regime, we conclude that oscillations are of the Coulomb type from the scaling of magnetic field period with the number of edges bound to the antidot. Based on both gate-voltage and field periods, we find at filling factor $\nu=2$ a tunneling charge of $e$ and two charged edges. Generalizing this picture to the fractional regime, we find (again, based on field and gate-voltage periods) at  $\nu=2/3$ a tunneling charge of $(2/3)e$ and a single charged edge. 

\end{abstract}

\maketitle

The fractional quantum Hall effect occurs when a high-mobility two-dimensional electronic gas (2DEG) is subject to a perpendicular applied magnetic field. At low temperature, electrons in the bulk of the 2DEG condense into incompressible states \cite{LaughlinPRL}, with extended states at the sample edge carrying charge, spin, and energy. It was theoretically shown that for Laughlin states such as filling factor $\nu=1/3$, a single chiral edge state carries excitations with fractional charge \cite{WenPRL}. A more complicated structure is predicted for hole-conjugate fractions such as $\nu = 2/3$, where counter-propagating edge states hybridize in the presence of edge disorder, leading to a forward-propagating charge mode and a reverse-propagating neutral mode \cite{JohnsonPRL, KaneFisher, KanePRL,MoorePRB}. Contrasts between the two counter-propagating modes at $\nu = 2/3$ and the simpler situation at $\nu = 2$, where two forward-propagating modes remain separated and independent, have been discussed theoretically \cite{KaneFisher, MoorePRB}.

Ashoori \textit{et al.}~investigated the $\nu=2/3$ edge experimentally using edge magnetoplasmon propagation and found only a single charged edge mode \cite{AshooriPRB}. Recent measurements of current noise through a quantum point contact (QPC) at bulk filling $\nu=2/3$ found finite shot noise at half transmission of the QPC, which was interpreted as indicating a single composite 2/3 edge \cite{BidPRL}. In addition, a tunneling charge of $e^*=(2/3)e$ was observed at low temperature, decreasing to $e/3$ above 0.1 K \cite{BidPRL}. Subsequent theory addressed these surprising results by considering the agglomeration of $e/3$ quasiparticles at low temperatures \cite{FerraroPRB}. 

\begin{figure}[b]
\center \label{figure1}
\includegraphics[width=3.375 in]{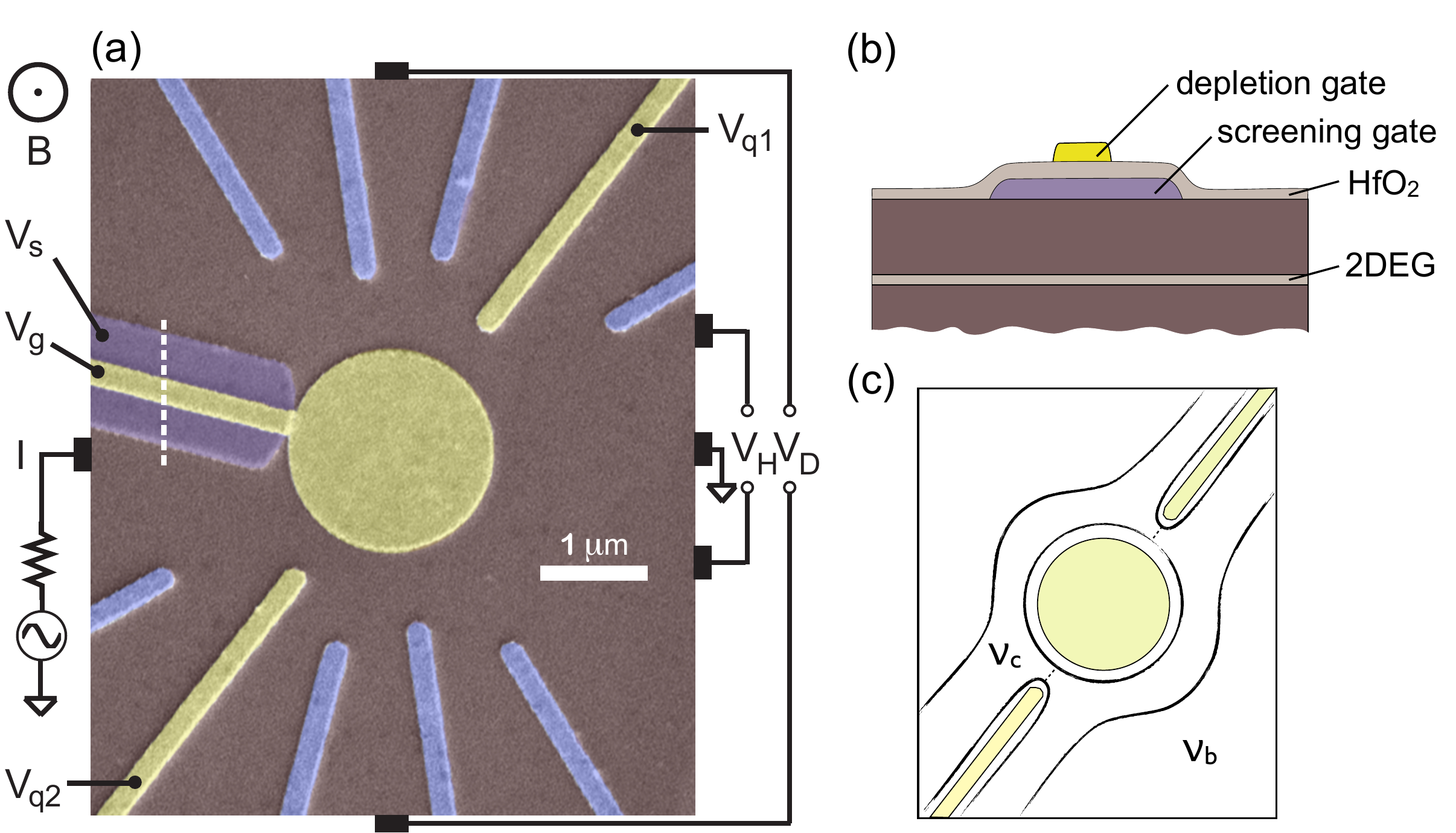}
\caption{\footnotesize{(a) False color SEM of the larger antidot device with $2~\um$ diameter. Diagonal voltage, $\vd$, across the device, as well as bulk Hall voltage, $V_H$, are measured as shown. Depleted gates are indicated in yellow, grounded gates are indicated in blue. The QPC constriction sizes range from 500~nm to 1.5~$\um$. (b) Cross section of the device taken along the dashed line in (a). The antidot gate is separated from the screening gate by a 30 nm $\mathrm{HfO_2}$ layer. (c) Schematic layout of the edge states through the device when $\vqpc=\vqpc^-$. The bulk filling factor, $\nu_b$, and the constriction filling factor, $\nu_c$, are indicated.}}
\end{figure}

Antidots have been used to study tunneling and confinement effects in both the integer and fractional quantum Hall regimes \cite{Hwang,SimpsonAPL,GoldmanSci,FordPRB,FranklinSurf,KataCoul,KataKondo,GoldmanPRB2001,GoldmanPRB2008,SimPR,Maasilta}. Early experiments by Hwang \textit{et al.}~\cite{Hwang} found resistance oscillations as a function of magnetic field and channel gate voltage for a density slightly below $\nu=1$ in the constrictions between the antidot and sample edges. The observed field period, corresponding to one flux quantum through the antidot, was interpreted in terms of a single-particle Aharonov-Bohm phase \cite{Hwang}. Subsequent experiments near $\nu=2$ in the constrictions found a field period corresponding to $h/2e$ ($\phi_0/2$) through the antidot, motivating an alternative interpretation in terms of Coulomb charging of two isolated edge states \cite{FordPRB}.  Coulomb charging of an antidot was observed directly by Kataoka \textit{et al.}~at both $\nu=1$ and $\nu=2$ using a QPC charge sensor \cite{KataCoul}. Goldman \textit{et al.} found the field period of oscillations to depend on the filling factor through the constriction, interpreting this dependence as a signature of Coulomb charging of multiple isolated edges surrounding the antidot \cite{GoldmanPRB2008}.  Recent theory by Ihnatsenka \textit{et al.}~captures many of the effects observed experimentally in antidots in the integer quantum Hall regime \cite{AdTheory}. Much less experimental work on antidots has been reported in the fractional regime despite numerous theoretical proposals related to this system \cite{Kivelson,Loss,DasSarma,Averin,Ady}. A key experiment was the measurement of fractional tunneling charge, $e^{*}=e/3$, for an antidot with $\nu=1/3$ in the constrictions, based on the back-gate voltage and magnetic field periods of observed resistance oscillations \cite{GoldmanSci}. 

In this Letter, we report measurements of resistance oscillations in gate-defined antidots of two sizes, comparing integer and fractional filling factors in the constrictions between the antidot and adjacent gates that extend to the sample edge. Oscillations as a function of perpendicular magnetic field and antidot gate voltage were measured in 2D sweeps, and dominant frequencies extracted from 2D Fourier spectra. For integer filling in the constrictions, magnetic field oscillation frequencies were found to be proportional to the filling factor in the constrictions, consistent with a Coulomb charging model. At $\nu=2/3$, the magnetic field oscillation frequency was found to be consistent with a single charged edge within a generalized Coulomb charging picture, with the charge-carrying edge state located slightly closer to the antidot than the single edge found at $\nu = 1$. Gate-voltage oscillations provide a direct measurement of the tunneling quasiparticle charge.  Normalizing to a tunneling charge of $e$ at $\nu = 1$, which determines the gate voltage lever arm, we find a tunneling charge consistent with $e$ for all measured integer filling factors, and a tunneling charge consistent with $e^*=(2/3)e$ at $\nu = 2/3$. 

Antidot devices with $1~\um$ and $2~\um$ diameter antidots were fabricated on a symmetrically Si-doped GaAs/AlGaAs 30~nm quantum well structure located 230~nm below the wafer surface, with density $n$~=~$1.6\times10^{15}~\perm$ and mobility $\mu$~=~1,200 $\mob$/V$\cdot$s measured in the dark. A Ti/Au (8 nm/42 nm) screening gate was first patterned using electron-beam lithography on a wet-etched mesa [purple gate in Fig.~1(a)]. The sample was then coated with 30~nm of HfO$_{2}$ using atomic layer deposition. A circular antidot [Ti/Au (8 nm/42 nm)] was next patterned on top of the HfO$_{2}$, positioned to extend beyond the edge of the screening gate, with connection to a remote bonding pad via a ``pan handle'' depletion gate that runs on top of the screening layer, so that only the 2DEG under the antidot is depleted when the gate is activated [Fig.~1(a, b)]. 

Transport measurements were made using a current bias $I$ of 0.3~nA at 101~Hz, with magnetic field, $B$, applied perpendicular to the plane of the 2DEG, in a dilution refrigerator with base temperature $\sim$10~mK. QPC gate voltages ($V_{\mathrm{q1}}, V_{\mathrm{q2}}$), were trimmed around $-1.0~\mathrm{V}$ to symmetrize the device, i.e., to give the same filling factors in the two constrictions. The antidot gate voltage, $\vg$, was then swept around $-0.9~\mathrm{V}$, yielding periodic resistance oscillations. The screening gate and all unused QPC gates were grounded. The diagonal voltage, $\vd$, was measured between incoming edge states on opposite sides of the device [Fig.~1(c)], and the diagonal resistance, $\rd \equiv d\vd/dI$, was used to determine the filling factor in the QPCs, $\vqpc=h/(\rd e^2)$. The bulk Hall resistance, $\rh \equiv d\vh/dI$, was simultaneously measured using contacts away from the antidot [Fig.~1(a)]. Figure 1(c) schematically shows the bulk filling factor, $\nu_b$, along with the filling factor in the constrictions, $\vqpc$, in the condition where $\rd$ is slightly larger than the resistance of well-quantized plateaus. This filling, measured on the high-field side of the $\vqpc$ plateau in $\rd$, is denoted $\vqpc^-$. 

\begin{figure}
\center \label{figure2}
\includegraphics[width=3.375 in]{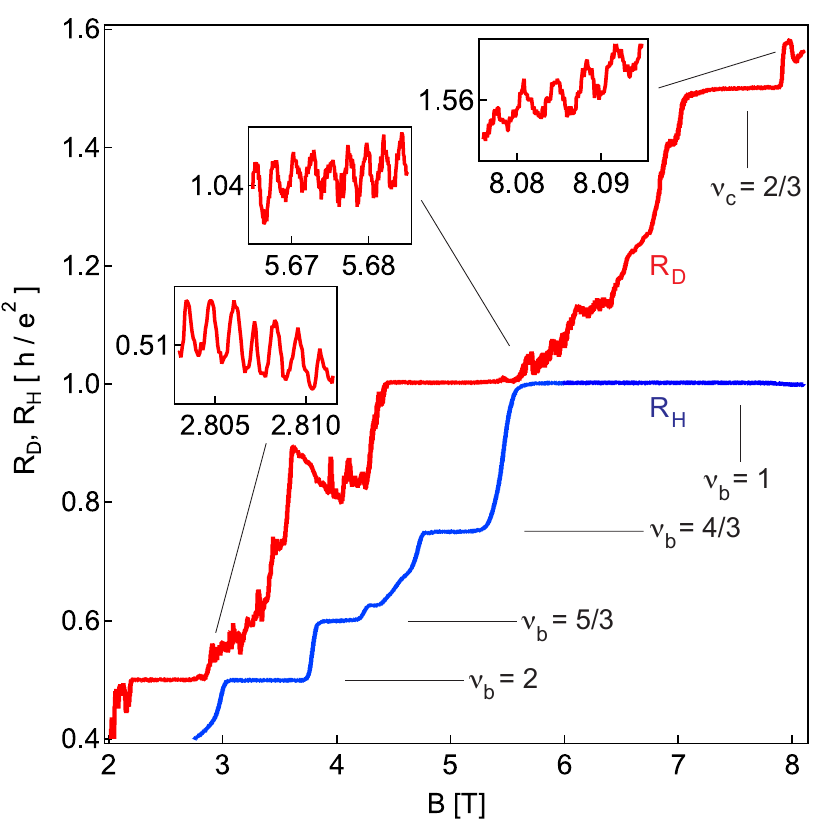}
\caption{\footnotesize{Diagonal resistance, $\rd$ (red), and bulk Hall resistance, $\rh$ (blue), as a function of perpendicular magnetic field, $B$, in the $1~\um$ diameter antidot with the 500 nm QPCs activated. Insets show regions of oscillations at $\vqpc=2^-$, $\vqpc=1^-$, and $\vqpc=2/3^-$.}}
\end{figure} 

Figure 2 shows $\rh(B)$ and $\rd(B)$ at fixed gate voltages in the $1~\um$ diameter antidot with the 500 nm QPCs activated. Zoom-ins reveal periodic oscillations in $\rd (B)$, with periods $\Delta B= 2.1~\mathrm{mT}$ for $\vqpc=1^-$, $\Delta B=1.0~\mathrm{mT}$ for $\vqpc=2^-$, and $\Delta B=2.9~\mathrm{mT}$ for $\vqpc=2/3^-$.  Aperiodic fluctuations in $\rd(B)$ were also observed on the low-field sides of the plateaus. The period of 2.1 mT at $\vqpc=1^-$ corresponds to $\phi_0=h/e$ through an area of 2.0$~\mathrm{\mu m}^2$, larger than the lithographic area of the antidot, $A=0.8~\mathrm{\mu m}^2$. We attribute the larger area to the finite depletion length of the antidot top gate, as discussed quantitatively below. The period of 1.0 mT at $\vqpc=2^-$ corresponds to $\phi_0/2$ going through a device of about the same area. Continuing this trend for integer states, oscillations observed at $\vqpc=3^-$ had a field period corresponding to $\phi_0/3$; oscillations at $\vqpc=4^-$ had a field period corresponding to $\phi_0/4$ (not shown). The observed scaling of field periods with constriction filling factor, i.e., the field period at $\vqpc$ corresponds to $\phi_0/\vqpc$, is consistent with previous experiments in the integer quantum Hall regime in antidots \cite{FordPRB,GoldmanPRB2008}.

\begin{figure}[t!]
\center \label{figure3}
\includegraphics[width=3.375 in]{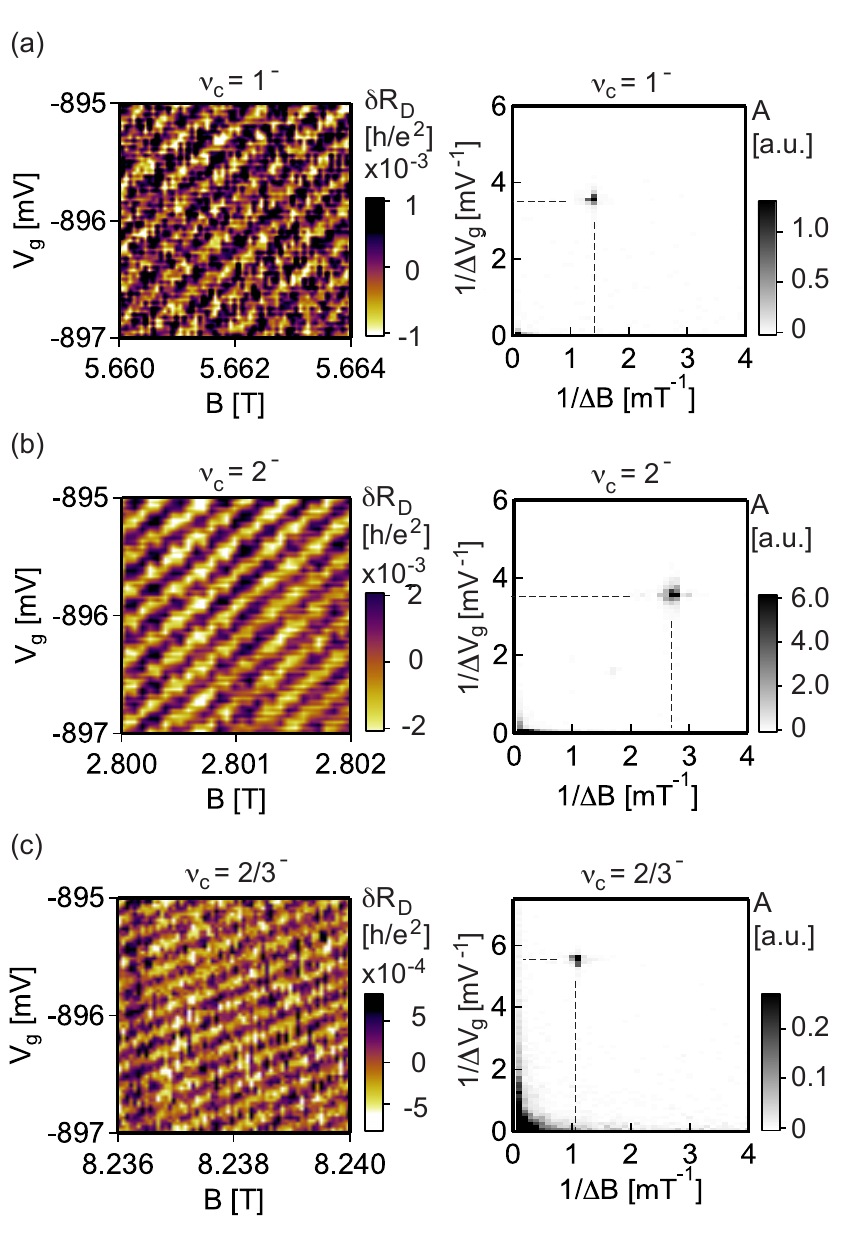}
\caption{\footnotesize{Diagonal resistance with background subtracted, $\rgd$, as a function of $B$ and $\vg$ in $2~\mathrm{\mu m}$ diameter antidot at (a) $\vqpc = 1^-$, (b) $\vqpc = 2^-$, and (c) $\vqpc = 2/3^-$, with corresponding 2D Fourier power spectra. Dominant peaks in the power spectra are marked by dashed lines, located at (1.4 mT$^{-1}$, 3.6 mV$^{-1}$) for $\vqpc=1^-$, (2.7 mT$^{-1}$, 3.6 mV$^{-1}$) for $\vqpc=2^-$, and (1.1 mT$^{-1}$, 5.6  mV$^{-1}$) for $\vqpc=2/3^-$.}}
\end{figure} 

Figure~3 shows $\rd$ with a smooth background subtracted, denoted $\rgd$, as a function of both $B$ and $\vg$ in the $2~\mathrm{\mu m}$ diameter antidot with 500 nm QPCs activated at $\vqpc=1^-$, $\vqpc=2^-$, and $\vqpc=2/3^-$ along with corresponding two-dimensional (2D) Fourier power spectra. The 2D plots of $\rgd$ reveal positively sloped stripes in all cases. The dominant peaks in the Fourier spectra show the expected scaling of the field period between $\vqpc=1^-$ and $\vqpc=2^-$, with peaks at $1.4~\mathrm{mT}^{-1}$ at $\vqpc=1^-$ and $2.7~\mathrm{mT}^{-1}$ at $\vqpc=2^-$ differing roughly by the same factor of two as discussed above. At $\vqpc=2/3^-$, the dominant Fourier peak is at $1.1~\mathrm{mT}^{-1}$, which is close to, but somewhat smaller than the magnetic field frequency at $\vqpc=1^-$. This scaling of magnetic field periods is similar to what was observed in the 1~$\mu$m device above.  The position of the dominant spectral peak as a function of gate-voltage frequency, $1/\Delta \vg$, is the same  for $\vqpc=1^-$ and $\vqpc=2^-$, in both cases $3.6~\mathrm{mV}^{-1}$, but increases to $5.6~\mathrm{mV}^{-1}$ at $\vqpc=2/3^-$.

Table I summarizes the observed oscillations as a function of magnetic field.  Magnetic field periods at each filling factor were used to find a best-fit effective area enclosed by the charged edge state, constrained to have the same depletion length, $d$, for two antidot sizes measured at the same antidot gate voltage. To calculate the depletion length from the data at $\vqpc=2^-$, we take one flux quantum to correspond to twice the measured magnetic field period. Note that the depletion length deduced in this way is $\sim 100$~nm smaller at $\vqpc=2/3^-$ than at $\vqpc=1^-$ and $\vqpc=2^-$. 

The number of charge-carrying edges, $N$, at each filling factor is deduced from the magnetic field period within a Coulomb charging picture, where each edge contributes one oscillation in resistance per flux quantum. In the integer regime, this interpretation is consistent with previous experiments and data in both antidots and quantum dots \cite{FordPRB, GoldmanPRB2008, Yiming, Nissim}, and reconciles the unphysically large effective area that would be inferred from the field period at $\vqpc=2^-$ from a single-edge picture. We find, therefore, $N=2$ separate charge-carrying edges bound to the antidot at $\vqpc=2^-$ . At $\vqpc=2/3^-$, the measured period is too large to correspond to two charged edges within an analogous Coulomb picture, as the large field period would imply $d<0$. We thus find $N =1$ charged edge at $\vqpc=2/3^-$. 

\begin{table}
\begin{ruledtabular}
\begin{tabular}{ccccc}
$D~[\mu$m] & $\nu_c$ & $d$ [nm] & $\Delta B$ [mT] & $N$\rule[-0.25cm]{0cm}{0.5cm}\\
\noalign{\hrule height 1pt}
\multirow{3}{*}{2} & 1 & 320 & 0.74 & 1\rule{0cm}{0.3cm}\\
 & 2 & 310 & 0.38 & 2 \\
 & 2/3 & 190 & 0.92 & 1 \\
 \hline
\multirow{3}{*}{1} & 1 & 320 & 2.1 & 1\rule{0cm}{0.3cm} \\
& 2 & 310 & 1.0 & 2 \\
 & 2/3 & 190 & 2.9 & 1 \\
\end{tabular}
\caption{Summary of magnetic field oscillations data, for antidots with lithographic diameter $D$ and  filling factor $\nu_{c}$ in the constrictions connecting the antidot to the sample edges. Oscillation periods, $\Delta B$, used to determine depletion length, $d$, taken to be the same for the two antidots at the same filling factor, and the number of edge channels, $N$. }
\end{ruledtabular}
\end{table}

\begin{table}
\begin{ruledtabular}
\begin{tabular}{cccc}
$D~[\mu$m] & $\nu_c$ & $\Delta V_{\rm g}$[mV] & $\frac{\Delta V_{\rm g}}{\Delta V_{\rm g}^{\nu_{c}=1}}$\rule[-0.375cm]{0cm}{0.75cm}
 \\
\noalign{\hrule height 1pt}
\multirow{3}{*}{2} & 1 & 0.28 & $\equiv$ 1\rule{0cm}{0.3cm}\\
 & 2 & 0.28 & 1.0 \\
 & 2/3 & 0.18 & 0.64 \\
 \hline
\multirow{3}{*}{1} & 1 & 0.62 & $\equiv$ 1\rule{0cm}{0.3cm} \\
& 2 & 0.62 & 1.0 \\
& 2/3 & 0.42 & 0.67 \\
\end{tabular}
\end{ruledtabular}
\caption{Summary of gate-voltage oscillations data, for antidots with lithographic diameter $D$ and filling factor $\nu_{c}$ in the constrictions. Oscillation periods, $\Delta V_{\rm g}$, and periods relative to the period at $\nu_{c} = 1^{-}$ determine the tunneling charge, $e^{*}$, as discussed in the text.}
\end{table}

Table II summarizes observed oscillations as a function of antidot gate voltage. Gate-voltage periods at $\vqpc=1^-$ and $\vqpc=2^-$ are the same within measurement uncertainty. The gate-voltage period at $\vqpc=2/3^-$, is $\sim 2/3$ of this value. No features are visible in the $\vqpc=2/3^-$ Fourier spectrum at 1/3 times the gate-voltage period at $\vqpc=1^-$ in either device. We also do not find any change in the gate-voltage period above 100~mK, in contrast to the behavior observed in Ref.~\cite{BidPRL}.

In summary, based on magnetic field and gate-voltage periods, we infer a Coulomb charging mechanism for the observed resistance oscillations. In the case of single-particle Aharonov-Bohm oscillations, one would expect a constant field period and a gate-voltage period that scales inversely with increasing applied magnetic field \cite{AdTheory}. We do not observe this behavior in our devices. Instead, we have found that the field period scales with the number of edges in the system and the gate-voltage period is constant for integer oscillations but changes for fractional oscillations. From these observations, we conclude that Coulomb effects are the dominant mechanisms for oscillations in our system \cite{LeePRL,AdTheory}. 

Coulomb oscillations in antidots in the integer regime was considered theoretically in Ref.~\cite{AdTheory}, which found that the field period scales with the number of fully transmitted edges in the constrictions. We instead find oscillations on the high field side of the $\vqpc$ plateau in $\rd$, where the outermost edge is not fully transmitted, but $\Delta \mathrm{B}$ still scales with $\vqpc$. Similar results were reported in Ref.~\cite{FordPRB}, which found a field period corresponding to $\phi_0/2$ on both sides of the $\nu=2$ plateau. We conclude that it is number of antidot-bound edge states that determines the magnetic field period.

The picture of Coulomb oscillations gives an antidot gate-voltage period proportional to the tunneling charge, $e^{*}$, independent of the number of edges \cite{LeePRL,AdTheory}, assuming a capacitive coupling (or lever arm) that is roughly independent of filling factor. Numerical modeling indicates that the capacitance of the antidot gate to a nearby 2DEG does not change significantly within the range of depletion lengths in Table I. The insensitivity of capacitance to filling factor is also supported by the observation of equal gate-voltage periods for $\vqpc=1^-$ and $\vqpc=2^-$.  The observed period at $\vqpc=2/3^-$ thus strongly suggests $e^{*}=2e/3$. This result is somewhat surprising in light of previous measurements of tunneling into a disorder-induced charge puddle at $\nu=2/3$, which found $e^{*}=e/3$ \cite{AYSci}. We speculate that the smaller charging energies in the current devices, due to screening from the antidot gate, allow a quasiparticle pairing energy associated with edge reconstruction at $\nu=2/3$ to dominate over the Coulomb energy associated with tunneling $2e/3$ rather than $e/3$. Further experiments, including on devices that allow direct charge sensing, and over a broad range of device areas, will help clarify this result.

We thank D.T. McClure, B. I. Halperin, M. Heiblum, B. Rosenow, and A. Yacoby for useful discussions, and P. Gallagher for experimental contributions. Research supported by Microsoft Project Q, the National Science Foundation (DMR-0501796), and the Department of Energy, Office of Science. Device fabrication used Harvard's Center for Nanoscale Systems.


\small
\begin{thebibliography}{31}

\expandafter\ifx\csname natexlab\endcsname\relax\def\natexlab#1{#1}\fi
\expandafter\ifx\csname bibnamefont\endcsname\relax
\def\bibnamefont#1{#1}\fi
\expandafter\ifx\csname bibfnamefont\endcsname\relax
\def\bibfnamefont#1{#1}\fi
\expandafter\ifx\csname citenamefont\endcsname\relax
\def\citenamefont#1{#1}\fi
\expandafter\ifx\csname url\endcsname\relax
\def\url#1{\texttt{#1}}\fi
\expandafter\ifx\csname urlprefix\endcsname\relax\def\urlprefix{URL }\fi
\providecommand{\bibinfo}[2]{#2} \providecommand{\eprint}[2][]{\url{#2}}

\bibitem{LaughlinPRL}
\bibinfo{author}{\bibfnamefont{R.~B.}~\bibnamefont{Laughlin}},
\bibinfo{journal}{Phys.~Rev.~Lett.} \textbf{\bibinfo{volume}{50,}} \bibinfo{pages}{1395} (\bibinfo{year}{1983}).

\bibitem{WenPRL}
\bibinfo{author}{\bibfnamefont{X.~G.}~\bibnamefont{Wen}},
\bibinfo{journal}{Phys.~Rev.~Lett.} \textbf{\bibinfo{volume}{64,}} \bibinfo{pages}{2206} (\bibinfo{year}{1990}).

\bibitem{JohnsonPRL}
\bibinfo{author}{\bibfnamefont{M.~D.}~\bibnamefont{Johnson}} and \bibinfo{author}{\bibfnamefont{A.~H.}~\bibnamefont{MacDonald}},
\bibinfo{journal}{Phys.~Rev.~Lett.} \textbf{\bibinfo{volume}{67,}} \bibinfo{pages}{2060} (\bibinfo{year}{1991}).


\bibitem{KaneFisher}
\bibinfo{author}{\bibfnamefont{C.~L.}~\bibnamefont{Kane}} and \bibinfo{author}{\bibfnamefont{M.~P.~A.}~\bibnamefont{Fisher}},
\bibinfo{journal}{Phys.~Rev.~B} \textbf{\bibinfo{volume}{51,}} \bibinfo{pages}{13449} (\bibinfo{year}{1994}).

\bibitem{KanePRL}
\bibinfo{author}{\bibfnamefont{C.~L.}~\bibnamefont{Kane}}, \bibinfo{author}{\bibfnamefont{M.~P.~A.}~\bibnamefont{Fisher}}, \bibinfo{author}{\bibfnamefont{J.}~\bibnamefont{Polchinski}},
\bibinfo{journal}{Phys.~Rev.~Lett.} \textbf{\bibinfo{volume}{72,}} \bibinfo{pages}{4129} (\bibinfo{year}{1994}).

\bibitem{MoorePRB}
\bibinfo{author}{\bibfnamefont{J.~E.}~\bibnamefont{Moore}} and \bibinfo{author}{\bibfnamefont{F.~D.~M.}~\bibnamefont{Haldane}},
\bibinfo{journal}{Phys.~Rev.~B} \textbf{\bibinfo{volume}{55,}} \bibinfo{pages}{7818} (\bibinfo{year}{1997}).

\bibitem{AshooriPRB}
\bibinfo{author}{\bibfnamefont{R.~C.}~\bibnamefont{Ashoori}} {\it et al.},
\bibinfo{journal}{Phys.~Rev.~B} \textbf{\bibinfo{volume}{45,}} \bibinfo{pages}{3894} (\bibinfo{year}{1992}).

\bibitem{BidPRL}
\bibinfo{author}{\bibfnamefont{A.}~\bibnamefont{Bid}} {\it et al.},
\bibinfo{journal}{Phys.~Rev.~Lett.} \textbf{\bibinfo{volume}{103,}} \bibinfo{pages}{236802} (\bibinfo{year}{2009}).

\bibitem{FerraroPRB}
\bibinfo{author}{\bibfnamefont{D.}~\bibnamefont{Ferraro}} {\it et al.},
\bibinfo{journal}{Phys.~Rev.~B} \textbf{\bibinfo{volume}{82,}} \bibinfo{pages}{085323} (\bibinfo{year}{2010}).

\bibitem{Hwang}
\bibinfo{author}{\bibfnamefont{S.~W.}~\bibnamefont{Hwang}} {\it et al.},
\bibinfo{journal}{Phys.~Rev.~B} \textbf{\bibinfo{volume}{44,}} \bibinfo{pages}{13497} (\bibinfo{year}{1991}).

\bibitem{SimpsonAPL}
\bibinfo{author}{\bibfnamefont{P.~J.}~\bibnamefont{Simpson}} {\it et al.},
\bibinfo{journal}{Appl.~Phys.~Lett.} \textbf{\bibinfo{volume}{63,}} \bibinfo{pages}{3191} (\bibinfo{year}{1993}).

\bibitem{FordPRB}
\bibinfo{author}{\bibfnamefont{C.~J.~B.}~\bibnamefont{Ford}} {\it et al.},
\bibinfo{journal}{Phys.~Rev.~B} \textbf{\bibinfo{volume}{49,}} \bibinfo{pages}{17456} (\bibinfo{year}{1994}).

\bibitem{GoldmanSci}
\bibinfo{author}{\bibfnamefont{V.~J.}~\bibnamefont{Goldman}} and \bibinfo{author}{\bibfnamefont{B.}~\bibnamefont{Su}}, \bibinfo{journal}{Science} \textbf{\bibinfo{volume}{267,}} \bibinfo{pages}{1010} (\bibinfo{year}{1995}).

\bibitem{FranklinSurf}
\bibinfo{author}{\bibfnamefont{J.~D.~F.}~\bibnamefont{Franklin}} {\it et al.},
\bibinfo{journal}{Surf.~Sci.} \textbf{\bibinfo{volume}{361/362,}} \bibinfo{pages}{17} (\bibinfo{year}{1996}).

\bibitem{Maasilta}
\bibinfo{author}{\bibfnamefont{I.~J.}~\bibnamefont{Maasilta}} and \bibinfo{author}{\bibfnamefont{V.~J.}~\bibnamefont{Goldman}},
\bibinfo{journal}{Phys.~Rev.~B} \textbf{\bibinfo{volume}{55,}} \bibinfo{pages}{4081} (\bibinfo{year}{1997}).

\bibitem{KataCoul}
\bibinfo{author}{\bibfnamefont{M.}~\bibnamefont{Kataoka}} {\it et al.},
\bibinfo{journal}{Phys.~Rev.~Lett.} \textbf{\bibinfo{volume}{83,}} \bibinfo{pages}{160} (\bibinfo{year}{1999}).

\bibitem{KataKondo}
\bibinfo{author}{\bibfnamefont{M.}~\bibnamefont{Kataoka}} {\it et al.},
\bibinfo{journal}{Phys.~Rev.~Lett.} \textbf{\bibinfo{volume}{89,}} \bibinfo{pages}{226803} (\bibinfo{year}{2002}).

\bibitem{GoldmanPRB2001}
\bibinfo{author}{\bibfnamefont{V.~J.}~\bibnamefont{Goldman}} {\it et al.},
\bibinfo{journal}{Phys.~Rev.~B} \textbf{\bibinfo{volume}{64,}} \bibinfo{pages}{085319} (\bibinfo{year}{2001}).

\bibitem{SimPR}
\bibinfo{author}{\bibfnamefont{H.-S.}~\bibnamefont{Sim}}, \bibinfo{author}{\bibfnamefont{M.}~\bibnamefont{Kataoka}}, \bibinfo{author}{\bibfnamefont{C.~J.~B.}~\bibnamefont{Ford}},
\bibinfo{journal}{Phys.~Rep.} \textbf{\bibinfo{volume}{456,}} \bibinfo{pages}{127} (\bibinfo{year}{2008}).

\bibitem{GoldmanPRB2008}
\bibinfo{author}{\bibfnamefont{V.~J.}~\bibnamefont{Goldman}}, \bibinfo{author}{\bibfnamefont{J.}~\bibnamefont{Liu}}, \bibinfo{author}{\bibfnamefont{A.}~\bibnamefont{Zaslavsky}},
\bibinfo{journal}{Phys.~Rev.~B} \textbf{\bibinfo{volume}{77,}} \bibinfo{pages}{115328} (\bibinfo{year}{2008}).

\bibitem{SimPRL}
\bibinfo{author}{\bibfnamefont{H.~S.}~\bibnamefont{Sim}} {\it et al.},
\bibinfo{journal}{Phys.~Rev.~Lett.} \textbf{\bibinfo{volume}{91,}} \bibinfo{pages}{266801} (\bibinfo{year}{2003}).

\bibitem{Jain}
\bibinfo{author}{\bibfnamefont{J.~K.}~\bibnamefont{Jain}} and \bibinfo{author}{\bibfnamefont{S.~A.}~\bibnamefont{Kivelson}},
\bibinfo{journal}{Phys.~Rev.~Lett.} \textbf{\bibinfo{volume}{60,}} \bibinfo{pages}{1542} (\bibinfo{year}{1988}).

\bibitem{Kivelson}
\bibinfo{author}{\bibfnamefont{S.~A.}~\bibnamefont{Kivelson}} and \bibinfo{author}{\bibfnamefont{V.~L.}~\bibnamefont{Pokrovsky}},
\bibinfo{journal}{Phys.~Rev.~B} \textbf{\bibinfo{volume}{40,}} \bibinfo{pages}{1373} (\bibinfo{year}{1989}).

\bibitem{Loss}
\bibinfo{author}{\bibfnamefont{M.~R.}~\bibnamefont{Geller}}, \bibinfo{author}{\bibfnamefont{D.}~\bibnamefont{Loss}}, \bibinfo{author}{\bibfnamefont{G.}~\bibnamefont{Kirczenow}},
\bibinfo{journal}{Phys.~Rev.~Lett.} \textbf{\bibinfo{volume}{77,}} \bibinfo{pages}{5110} (\bibinfo{year}{1996}).

\bibitem{DasSarma}
\bibinfo{author}{\bibfnamefont{S.}~\bibnamefont{Das~Sarma}}, \bibinfo{author}{\bibfnamefont{M.}~\bibnamefont{Freedman}}, \bibinfo{author}{\bibfnamefont{C.}~\bibnamefont{Nayak}},
\bibinfo{journal}{Phys.~Rev.~Lett.} \textbf{\bibinfo{volume}{94,}} \bibinfo{pages}{166802} (\bibinfo{year}{2005}).

\bibitem{Averin}
\bibinfo{author}{\bibfnamefont{D.~V.}~\bibnamefont{Averin}} and \bibinfo{author}{\bibfnamefont{J.~A.}~\bibnamefont{Nesteroff}},
\bibinfo{journal}{Phys.~Rev.~Lett.} \textbf{\bibinfo{volume}{99,}} \bibinfo{pages}{096801} (\bibinfo{year}{2007}).

\bibitem{Ady}
\bibinfo{author}{\bibfnamefont{R.}~\bibnamefont{Ilan}}, \bibinfo{author}{\bibfnamefont{B.}~\bibnamefont{Rosenow}}, \bibinfo{author}{\bibfnamefont{A.}~\bibnamefont{Stern}},
\bibinfo{journal}{Phys.~Rev.~Lett.} \textbf{\bibinfo{volume}{106,}} \bibinfo{pages}{136801} (\bibinfo{year}{2011}).

\bibitem{AdTheory}
\bibinfo{author}{\bibfnamefont{S.}~\bibnamefont{Ihnatsenka}}, \bibinfo{author}{\bibfnamefont{I.~V.}~\bibnamefont{Zozoulenko}}, \bibinfo{author}{\bibfnamefont{G.}~\bibnamefont{Kirczenow}},
\bibinfo{journal}{Phys.~Rev.~B} \textbf{\bibinfo{volume}{80,}} \bibinfo{pages}{115303} (\bibinfo{year}{2009}).

\bibitem{Yiming}
\bibinfo{author}{\bibfnamefont{Y.}~\bibnamefont{Zhang}} {\it et al.},
\bibinfo{journal}{Phys.~Rev.~B} \textbf{\bibinfo{volume}{79,}} \bibinfo{pages}{241304(R)} (\bibinfo{year}{2009}).

\bibitem{Nissim}
\bibinfo{author}{\bibfnamefont{N.}~\bibnamefont{Ofek}} {\it et al.},
\bibinfo{journal}{Proc.~Natl.~Acad.~Sci.~USA} \textbf{\bibinfo{volume}{107,}} \bibinfo{pages}{5276} (\bibinfo{year}{2010}).


\bibitem{LeePRL}
\bibinfo{author}{\bibfnamefont{P.~A.}~\bibnamefont{Lee}},
\bibinfo{journal}{Phys.~Rev.~Lett.} \textbf{\bibinfo{volume}{65,}} \bibinfo{pages}{2206} (\bibinfo{year}{1990}).

\bibitem{LeePRB}
\bibinfo{author}{\bibfnamefont{W.~R.}~\bibnamefont{Lee}} and \bibinfo{author}{\bibfnamefont{H.~S.}~\bibnamefont{Sim}},
\bibinfo{journal}{Phys.~Rev.~B} \textbf{\bibinfo{volume}{83,}} \bibinfo{pages}{035308} (\bibinfo{year}{2011}).


\bibitem{AYSci}
\bibinfo{author}{\bibfnamefont{J.}~\bibnamefont{Martin}} {\it et al.},
\bibinfo{journal}{Science} \textbf{\bibinfo{volume}{305,}} \bibinfo{pages}{980} (\bibinfo{year}{2004}).


 \end{thebibliography}
\end{document}